\documentclass[thmsa,letterpaper]{article}

\usepackage{amssymb}

\begin{document}

\author{David Carf\`{i}}
\title{Reactivity in decision-form games}
\date{}
\maketitle

\begin{abstract}
In this paper we introduce the reactivity in decision-form games. The
concept of reactivity allows us to give a natural concept of rationalizable
solution for decision-form games: the solubility by elimination of
sub-reactive strategies. This concept of solubility is less demanding than
the concept of solubility by elimination of non-reactive strategies
(introduced by the author and already studied and applied to economic
games). In the work we define the concept of super-reactivity, the preorder
of reactivity and, after a characterization of super-reactivity, we are
induced to give the concepts of maximal-reactivity and sub-reactivity; the
latter definition permits to introduce the iterated elimination of
sub-reactive strategies and the solubility of a decision-form game by
iterated elimination of sub-reactive strategies. In the paper several
examples are developed. Moreover, in the case of normal-form games, the
relation between reactivity and dominance is completely revealed.
\end{abstract}

\bigskip

\section{\textbf{Introduction}}

\bigskip

In this paper we introduce the concept of \emph{reactivity for two-player
decision-form games} and concentrate upon it. For the concept of decision
form game, formally introduced and developed by the author himself, the
reader can see \cite{CA2} and \cite{CA3}; for the origin of the concept and
its motivation the reader can see \cite{AU1} and \cite{AU2}. Let $G=(e,f)$
be a decision-form game and let us christen our two player Emil and Frances,
it is quite natural that if an Emil's strategy $x$ can react to all the
Frances' strategies to which an other strategy $x^{\prime }$ can react, then
we must consider the strategy $x$ \emph{reactive} at least as the strategy $%
x^{\prime }$; moreover, if the strategy $x$ is reactive at least as $%
x^{\prime }$ and $x$ can react to a Frances' strategy to which $x^{\prime }$
cannot react to, then Emil should consider the strategy $x$ strictly more
reactive than $x^{\prime }$. The previous simple considerations allow us to
introduce the \emph{capacity of reaction}, or reactivity, of any Emil's
strategy and to compare it with the capacity of reaction of the other Emil's
strategies. In this direction, we introduce the \emph{super-reactive
strategies} of a player $i$, i.e. strategies of player $i$ capable to reply
to any opponent's actions to which the player $i$ can reply: obviously these
strategies (whenever they there exist) are the best ones to use, in the
sense explained before. In a second time, we introduce the \emph{reactivity
comparison} between strategies and we observe that this relation is a
preorder. Then, we define the concept of reactivity and explain the nature
of the super-reactivity, this permits to define the concepts of \emph{%
maximally reactive strategy}, \emph{minimally reactive strategy, }and of 
\emph{sub-reactive strategy}. The concept of sub-reactivity will give us the
opportunity to introduce the principal operative concepts of the paper, i.e. 
\emph{the elimination of sub-reactive strategies}, the concept of \emph{%
reducing sequence of a game by elimination of sub-reactive strategies} and,
at last, the \emph{solvability of a game by elimination of sub-reactive
strategies} with the meaning of solution in the case of solvability.

\bigskip

\section{\textbf{Super-reactive strategies}}

\bigskip

\textbf{Definition (of super-reactive strategy).} \emph{Let }$(e,f)$\emph{\
be a two player decision-form game. An Emil's strategy }$x_{0}$\emph{\ is
called \textbf{super-reactive with respect to the decision rule }}$e$\emph{\
if it is a possible reaction to all the Frances' strategies to which Emil
can react. In other terms, an Emil's strategy }$x_{0}$\emph{\ is called
super-reactive if it belongs to the reaction set }$e(y)$\emph{, for each
Frances' strategy }$y$\emph{\ belonging to the domain of the decision rule }$%
e$\emph{. Analogously, a Frances' strategy }$y_{0}$\emph{\ is called \textbf{%
\ \ super-reactive with respect to the decision rule }}$f$\emph{\ if it is a
possible reaction to all the Emil's strategies to which Frances can react.
In other terms, a Frances' strategy }$y_{0}$\emph{\ is called super-reactive
if it belongs to the reaction set }$f(x)$\emph{, for each Emil's strategy }$%
x $\emph{\ in the domain of the decision rule }$f$\emph{.}

\bigskip

\textbf{Remark.} Let $E^{\prime }$ be the domain of the decision rule $f$
and $F^{\prime }$ be the domain of the decision rule $e$. The sets of all
the Frances' and Emil's super-reactive strategies are the two intersections 
\[
\cap ^{\neq }(e):=\cap _{y\in F^{\prime }}e(y)
\]
and
\[
\cap ^{\neq }(f):=\cap _{x\in E^{\prime }}f(x),
\]
respectively. If Frances has no disarming strategies toward Emil we have 
\[
\cap ^{\neq }(e)=\cap e:=\cap _{y\in F}e(y).
\]
Analogously, if Emil has no disarming strategies toward Frances 
\[
\cap ^{\neq }(f)=\cap f:=\cap _{x\in E}f(x).
\]
Obviously these two intersections can be empty.

\bigskip

We note here an elementary and obvious result.

\bigskip

\textbf{Proposition.} \emph{Let }$(e,f)$\emph{\ be a decision form game and
let }$x_{0}$\emph{\ and }$y_{0}$\emph{\ be two non-disarming and
super-reactive strategies of the first and second player respectively. Then
the bistrategy }$(x_{0},y_{0})$\emph{\ is an equilibrium of the game.}

\bigskip

It is straightforward that a game can have equilibria and lack in
super-reactive strategy, as the following example shows.

\bigskip

\textbf{Example (of game without super-reactive strategies). }Let $(e,f)$ be
the decision form game with strategy spaces $E=\left[ -1,2\right] $ and $%
F=\left[ -1,1\right] $ and decision rules $e:F\rightarrow E$ and $%
f:E\rightarrow F$ defined by 
\begin{eqnarray*}
e(y) &=&\left\{ 
\begin{array}{cl}
-1 & \mathrm{if}\;y<0 \\ 
E & \mathrm{if}\;y=0 \\ 
2 & \mathrm{if}\;y>0
\end{array}
\right. , \\
f(x) &=&\left\{ 
\begin{array}{cl}
-1 & \mathrm{if}\;x<1 \\ 
F & \mathrm{if}\;x=1 \\ 
1 & \mathrm{if}\;x>1
\end{array}
\right. .
\end{eqnarray*}
Emil has not super-reactive strategies, in fact 
\[
\cap e=\left\{ -1\right\} \cap E\cap \left\{ 2\right\} =\varnothing .
\]
Also Frances has no super-reactive strategies, in fact 
\[
\cap f=\left\{ -1\right\} \cap F\cap \left\{ 1\right\} =\varnothing .
\]
Note that this game has three equilibria.

\bigskip

We say that an equilibrium of a game is a \emph{super-reactive equilibrium}
when it is a super-reactive cross, i.e. when it is a pair of super-reactive
strategies.

\bigskip

\textbf{Example (of game with super-reactive strategies). }Let $(e,f)$ be
the game with strategy spaces $E=\left[ -1,2\right] $ and $F=\left[
-1,1\right] $ and decision rules $e:F\rightarrow E$ and $f:E\rightarrow F$
defined by 
\begin{eqnarray*}
e(y) &=&\left\{ 
\begin{array}{cl}
\left[ -1,1\right]  & \mathrm{if}\;y<0 \\ 
E & \mathrm{if}\;y=0 \\ 
\left[ 0,2\right]  & \mathrm{if}\;y>0
\end{array}
\right. , \\
f(x) &=&\left\{ 
\begin{array}{cl}
-1 & \mathrm{if}\;x<1 \\ 
F & \mathrm{if}\;x=1 \\ 
\left\{ -1,1\right\}  & \mathrm{if}\;x>1
\end{array}
\right. .
\end{eqnarray*}
Emil has infinite super-reactive strategies, in fact the intersection of the
family (correspondence) $e$ is
\[
\cap e=\left[ -1,1\right] \cap E\cap \left[ 0,2\right] =\left[ 0,1\right] ,
\]
all the strategies $x$ between $0$ and $1$ are super-reactive for Emil.
Frances has only one super-reactive strategy, in fact 
\[
\cap f=\left\{ -1\right\} \cap F\cap \left\{ -1,1\right\} =\left\{
-1\right\} .
\]
Note that this game has infinitely many equilibria, their set is the graph
of the correspondence $f_{1}:E\rightarrow F$ defined by 
\[
f_{1}(x)=\left\{ 
\begin{array}{cl}
-1 & \mathrm{if}\;x<1 \\ 
F & \mathrm{if}\;x=1 \\ 
1 & \mathrm{if}\;x>1
\end{array}
\right. .
\]
On the other hand, only the equilibria belonging to the segment $\left[
0,1\right] \times \left\{ -1\right\} $ are super-reactive equilibria. Thanks
to super-reactivity, in this game an equilibrium is non-cooperatively
reachable; indeed, it is reasonable for Frances to play his unique
super-reactive strategy $-1$ and for Emil to play one of his super-reactive
strategies $x$ in $\left[ 0,1\right] $, consequently the game finishes in
the equilibrium $(x,-1)$.

\bigskip

\textbf{Remark (independence of the super-reactivity on the rival's rule).}
The Emil's (Frances's) super-reactive strategies depend only upon the Emil's
(Frances's) decision rule, and not on both the decision rules.

\bigskip

\textbf{Example (game with super-reactive strategies). }Let $E$ be the
compact interval $\left[ 0,1\right] $ and let $F$ be the interval $\left[
-1,1\right] $, let $e:F\rightarrow E$ be the correspondence defined by $%
e(y)=\left[ 0,\left| y\right| \right] $, for each $y$ in $F$. Frances has no
disarming strategies toward Emil. The strategy $0$ is the only Emil's
super-reactive strategy, because 
\[
\cap e=\bigcap_{y\in F}\left[ 0,\left| y\right| \right] =\left\{ 0\right\} . 
\]
Let $f:E\rightarrow F$ be defined by $f(x)=\left[ -x,x\right] $. Emil has no
disarming strategies toward Frances. The strategy $0$ is the only Frances'
super-reactive strategy, because 
\[
\cap f=\bigcap_{x\in E}\left[ -x,x\right] =\left\{ 0\right\} . 
\]
In this case we have again infinitely many equilibria, the points of the
graph of the correspondence $f_{1}:E\rightarrow F$ defined by $%
f_{1}(x)=\left\{ -x,x\right\} $, but we have only one super-reactive
equilibrium: the strategy profile $(0,0)$.

\bigskip

\bigskip

\section{\textbf{Comparison of reactivity}}

\bigskip

The definition of super-reactive strategy can be generalized.

\bigskip

\textbf{Definition (of comparison among reactivity).} \emph{Let }$(e,f)$%
\emph{\ be a two player decision form game. Let }$x_{0}$\emph{\ and }$x$%
\emph{\ be two Emil's strategies. We say that the strategy }$x_{0}$\emph{\
is \textbf{more reactive (in wide sense), with respect to the decision rule} 
}$e$\emph{, \textbf{than the strategy }}$x$\emph{, and we write }$x_{0}\geq
_{e}x$\emph{, if }$x_{0}$\emph{\ is a possible reaction to all the Frances'
strategies to which }$x$\emph{\ can react. In other terms, an Emil's
strategy }$x_{0}$\emph{\ is said more reactive than an other strategy }$x$%
\emph{\ when }$x_{0}$\emph{\ belongs to the reaction set }$e(y)$\emph{, for
each strategy }$y\in e^{-}(x)$\emph{.\ Analogously, let }$y_{0}$\emph{\ and }%
$y$\emph{\ be two Frances' strategies. We say that }$y_{0}$\emph{\ is more%
\textbf{\ reactive, with respect to the decision rule }}$f$\emph{, \textbf{\
than the strategy }}$y$\emph{, and we write }$y_{0}\geq _{f}y$\emph{, if the
strategy }$y_{0}$\emph{\ is a possible reaction to all the Emil's strategies
to which }$y$\emph{\ is a possible reaction. In other terms, a Frances'
strategy }$y_{0}$\emph{\ is more reactive than }$y$\emph{\ when }$y_{0}$%
\emph{\ belongs to the reaction set }$f(x)$\emph{, for each strategy }$x\in
f^{-}(y)$\emph{.}

\bigskip

\textbf{Memento (reciprocal correspondence). }We remember that the set $%
e^{-}(x)$ is the set of those Frances' strategies to which the strategy $x$
can reply, with respect to the decision rule $e$. In fact, the reciprocal
image of the strategy $x$ with respect to the correspondence $e$ is 
\[
e^{-}(x)=\left\{ y\in F:x\in e(y)\right\} , 
\]
therefore it is defined, exactly, as the set of all those Frances'
strategies $y$ for which $x$ is a possible response strategy. The reciprocal
correspondence of $e$, i.e. the correspondence $e^{-}:E\rightarrow F$
defined by $x\mapsto e^{-}(x)$, associates with every Emil's strategy $x$
the set of all those Frances's strategies for which $x$ is a possible
reaction. This last circumstance explains the interest in the determination
of the reciprocal correspondence $e^{-}$.

\bigskip

\textbf{Example (of comparison of reactivity). }Let $(e,f)$ be the decision
form game with strategy spaces $E=\left[ -1,2\right] $ and $F=\left[
-1,1\right] $ and decision rules $e:F\rightarrow E$ and $f:E\rightarrow F$
defined by 
\begin{eqnarray*}
e(y) &=&\left\{ 
\begin{array}{cl}
\left\{ -1\right\}  & \mathrm{if}\;y<0 \\ 
E & \mathrm{if}\;y=0 \\ 
\left\{ 2\right\}  & \mathrm{if}\;y>0
\end{array}
\right. , \\
f(x) &=&\left\{ 
\begin{array}{cl}
\left\{ -1\right\}  & \mathrm{if}\;x<1 \\ 
F & \mathrm{if}\;x=1 \\ 
\left\{ 1\right\}  & \mathrm{if}\;x>1
\end{array}
\right. .
\end{eqnarray*}
we want to determine the reciprocal multifunctions of $e$ and $f$. We have 
\begin{eqnarray*}
e^{-}(x) &=&\left\{ 
\begin{array}{cl}
\left[ -1,0\right]  & \mathrm{if}\;x=-1 \\ 
\left\{ 0\right\}  & \mathrm{if}\;x\in \left] -1,2\right[  \\ 
\left[ 0,1\right]  & \mathrm{if}\;x=2
\end{array}
\right. , \\
f^{-}(y) &=&\left\{ 
\begin{array}{cl}
\left[ -1,1\right]  & \mathrm{if}\;y=-1 \\ 
\left\{ 1\right\}  & \mathrm{if}\;y\in \left] -1,1\right[  \\ 
\left[ 1,2\right]  & \mathrm{if}\;y=1
\end{array}
\right. .
\end{eqnarray*}
From these two relations, we can easily observe that Emil's strategies $-1$
and $2$ are more reactive than all the other Emil's strategies of the
interval $\left] -1,2\right[ $, with respect to the rule $e$. To this aim,
we have to prove that the strategies $1$ and $2$\ belong to the reaction set 
$e(y)$, for each strategy $y\in e^{-}(x)$. Let $x$ be an Emil's strategy in
the open interval $\left] -1,2\right[ $, we have $e^{-}(x)=\left\{ 0\right\} 
$, then the relation $y\in e^{-}(x)$ is equivalent to $y=0$, but the image $%
e(0)$ is the whole of $E$, therefore it includes both $-1$ and $2$.
Analogously, we prove that Frances' strategies $-1$ and $1$ are more
reactive than every other strategy $y\in \left] -1,1\right[ $, with respect
to the decision rule $f$.

\bigskip

The following theorem expresses the reactivity comparison in a conditional
form.

\bigskip

\textbf{Theorem.}\emph{\ In the conditions of the above definition. An
Emil's strategy }$x_{0}$\emph{\ is more reactive than another Emil's
strategy }$x$\emph{, with respect to the decision rule }$e$\emph{, if, for
each Frances' strategy }$y$\emph{, from the relation }$x\in e(y)$\emph{\ it
follows }$x_{0}\in e(y)$\emph{. In symbols, the relation }$x_{0}\geq _{e}x$%
\emph{\ holds if and only if } 
\[
(\forall y\in F)(x\in e(y)\Rightarrow x_{0}\in e(y))\emph{.} 
\]
\emph{Analogously, a Frances' strategy }$y_{0}$\emph{\ is more reactive than
another Frances' strategy }$y$\emph{, with respect to the decision rule }$f$%
\emph{, if for each Emil's strategy }$x$\emph{, from the relation }$y\in
f(x) $\emph{\ we deduce }$y_{0}\in f(x)$\emph{. In symbols, the relation }$%
y_{0}\geq _{f}y$\emph{\ holds if and only if } 
\[
(\forall x\in E)(y\in f(x)\Rightarrow y_{0}\in f(x))\emph{.} 
\]

\bigskip

\bigskip

\section{\textbf{The reactivity preorder}}

\bigskip

It is immediate to verify that the relation of reactivity comparison
determined by the decision rule $f$ upon the strategy space $F$ - defined,
for each pair of strategies $(y,y^{\prime })$, by $y\geq _{f}y^{\prime }$,
and that we denote by $\geq _{f}$ - is a preorder. This justifies the
following definition.

\bigskip

\textbf{Definition (of reactivity preorder). }\emph{Let }$(e,f)$ \emph{be} 
\emph{a decision form game upon the underlying strategy pair }$(E,F)$\emph{.
The binary relation }$\geq _{f}$\emph{\ on the strategy set }$F$ \emph{is
called \textbf{preorder of reactivity induced by the \emph{\textbf{decision} 
}rule }}$f$\emph{\ \textbf{on\emph{\emph{\ Frances}}' strategy space}.
Symmetrically, the binary relation }$\geq _{e}$\emph{\ on the strategy space 
}$E$ \emph{is called \textbf{preorder of reactivity induced by the \emph{%
\textbf{decision} }rule }}$e$\emph{\ \textbf{on Emil's \emph{strategy }space}
.}

\emph{\bigskip }

\textbf{Remark (strict preorder of reactivity). }Since the reactivity
comparison $\geq _{f}$ is a preorder, it has an associated strict preorder:
the preorder $>_{f}$ defined, as usual in Preorder Theory, for each pair of
strategies $(y_{0},y)$, by $y_{0}>_{f}y$ if and only if $y_{0}\geq y$ and $%
y\ngeq y_{0}$. Analogous consideration holds for Emil.

\bigskip

Now we see an example of strict comparison of reactivity among strategies.

\bigskip

\textbf{Example (of strict reactivity comparison). }Let $(e,f)$ be the
decision form game with strategy spaces $E=\left[ -1,2\right] $ and $%
F=\left[ -1,1\right] $ and with decision rules $e:F\rightarrow E$ and $%
f:E\rightarrow F$ defined by 
\begin{eqnarray*}
e(y) &=&\left\{ 
\begin{array}{cl}
-1 & \mathrm{if}\;y<0 \\ 
E & \mathrm{if}\;y=0 \\ 
2 & \mathrm{if}\;y>0
\end{array}
\right. , \\
f(x) &=&\left\{ 
\begin{array}{cl}
-1 & \mathrm{if}\;x<1 \\ 
F & \mathrm{if}\;x=1 \\ 
1 & \mathrm{if}\;x>1
\end{array}
\right. .
\end{eqnarray*}
We apply the conditional characterization to prove (again) that Emil's
strategies $-1$ and $2$ are more reactive than all the strategies of the
open interval $\left] -1,2\right[ $. In fact, for each Frances' strategy $y$
, if an Emil's strategy $x\in \left] -1,2\right[ $ belongs to the reaction
set $e(y)$ of $y$, the Frances' strategy $y$ must necessarily be $0$
(because the unique reaction set $e(y)$ containing strategies different from 
$-1$ and $2$ is just $e(0)$), but, in this case, we have also that $-1$ and $
2$ belong to $e(y)$ (inasmuch, the reaction set $e(0)$ is the whole of $E$).
We have so proved that the inequality $-1,2\geq _{e}x$ holds for each
strategy $x$ in $E$. Now we want to prove that the strict inequality $%
-1,2>_{e}x$ holds true, for each $x\in \left] -1,2\right[ $ (i.e. that the
strategies $-1$ and $2$ are strictly more reactive than any other Emil's
strategy). It is sufficient to prove that, for instance, the relation $2\leq %
_{e}x$ is false, for any $x$ in $\left] -1,2\right[ $; for, fixed such an $x$%
, we have to show that there exists a strategy $y$ in $F$ such that $2\in
e(y)$ and $x\notin e(y)$ (i.e. a strategy $y$ in $F$ to which $2$ reacts and 
$x$ does not). Let $y=1$, we have $e(y)=\left\{ 2\right\} $, then $2$ is in $%
e(y)$ and any $x\in \left] -1,2\right[ $ does not.

\bigskip

\section{\textbf{The reactivity of a strategy}}

\bigskip

\textbf{Terminology (reciprocal decision rule).} Let $f:E\rightarrow F$ be a
Frances' decision rule. We can associate, in a natural way, with the
correspondence $f$ the Emil's decision rule 
\[
f^{-}:F\rightarrow E:y\mapsto f^{-}(y), 
\]
that we call \emph{Emil's decision rule reciprocal of the Frances's decision
rule }$f$. This reciprocal decision rule is canonically associated with the
application of $F$ into the set of subsets of $E$ associating with every
Frances's strategy $y$ the set of all Emil's strategies for which $y$ is a
possible reaction: the function 
\[
f^{\leftarrow }:F\rightarrow \mathcal{P}(E):y\mapsto f^{-}(y). 
\]
With abuse of language, we will name this application \emph{reciprocal
function of the correspondence} $f$.

\bigskip

\textbf{Theorem (characterization of the preorder of reactivity).}\emph{\
The (opposite) reactivity preorder }$\leq _{f}$\emph{\ is the preorder
induced (in the usual sense) by the reciprocal function of the decision rule 
}$f$\emph{, that is by the function} 
\[
f^{\leftarrow }:F\rightarrow \mathcal{P}(E):y\mapsto f^{-}(y), 
\]
\emph{endowing the set of parts of }$E$\emph{\ (denoted by }$\mathcal{P}(E)$%
\emph{) with the set inclusion order }$\subseteq $\emph{. In other terms,
the (opposite) preorder of reactivity }$\leq _{f}$\emph{\ is the reciprocal
image of the set inclusion order with respect to the reciprocal function }$%
f^{\leftarrow }$\emph{\ of the decision rule }$f$\emph{.}

\emph{\bigskip }

\emph{Proof.} Let $x\in E$ and $y\in F$ be strategies. The relation $y\in
f(x)$ is equivalent to the relation $x\in f^{-}(y)$, therefore a Frances's
strategy $y_{0}$ is more reactive than $y$ if and only if $f^{-}(y)\subseteq
f^{-}(y_{0})$. $\blacksquare $

\bigskip

The above characterization allows to give the following definition.

\bigskip

\textbf{Definition (of reactivity).}\emph{\ Let }$(e,f)$\emph{\ be a
decision form game. For each strategy }$x$\emph{\ in }$E$\emph{, the
reciprocal image of the strategy }$x$\emph{\ by the correspondence }$e$\emph{%
\ , that is the set }$e^{-}(x)$\emph{, is called the \textbf{reactivity of} }%
$x $\emph{\ with respect to the decision rule }$e$\emph{. Analogously, for
each France' strategy }$y$\emph{\ in }$F$\emph{, the reciprocal image of }$y$%
\emph{\ by the decision rule }$f$\emph{\ is called the \textbf{reactivity of}
}$y$\emph{\ with respect to }$f$.

\bigskip

\section{\textbf{Super-reactive strategies as maxima}}

\bigskip

The following obvious result characterizes super-reactive strategies of a
player as maxima (upper optima) of the strategy space of the player with
respect to the reactivity preorder induced by his decision rule. Therefore
it allows to reduce the concept of super-reactive strategy to the concept of
optimum.

\bigskip

\textbf{Theorem (characterization of super-reactivity).}\emph{\ Let }$(e,f)$ 
\emph{be} \emph{a decision form game. Any Frances' super-reactive strategy
is a maximum of the preorder space }$(F,\geq _{f})$\emph{\ and vice versa.}

\bigskip

\textbf{Remark (on the nature of super-reactive strategies).} After the
realization of the true nature of super-reactive strategies, we can observe
some of the previous examples in another way. We have, in fact, seen that
there are situations in which Frances has no super-reactive strategies, this
simply means that the preordered space $(F,\geq _{f})$ has no maxima; this
does not surprise, in fact a preordered space has maxima only in very
particular cases. Obviously, when a space has no maxima (as observed in
preordered space theory) we have to look for other solutions of the
corresponding decision problem (Pareto boundaries, cofinal and coinitial
parts, suprema and infima, and so on) but we shall analyze these aspects in
the following paragraphs.

\bigskip

We should notice that, in general, the space $(F,\geq _{f})$ is not an
ordered space, and therefore several maxima can exist (they must necessarily
be indifferent between themselves by the theorem of indifference of optima
in preordered spaces), as the following example shows.

\bigskip

\textbf{Example (of distinct super-reactive strategies). }Let $(e,f)$ be the
decision form game with strategy spaces the two intervals of the real line $%
E=\left[ a,b\right] $ and $F=\left[ c,d\right] $ and with decision rules $%
e:F\rightarrow E$ and $f:E\rightarrow F$ defined by 
\begin{eqnarray*}
e(y) &=&\left\{ 
\begin{array}{cl}
\left\{ a,b\right\}  & \mathrm{if}\;y<0 \\ 
E & \mathrm{if}\;y=0 \\ 
\left\{ a,b\right\}  & \mathrm{if}\;y>0
\end{array}
\right. , \\
f(x) &=&\left\{ 
\begin{array}{cl}
\left\{ c,d\right\}  & \mathrm{if}\;x<1 \\ 
F & \mathrm{if}\;x=1 \\ 
\left\{ c,d\right\}  & \mathrm{if}\;x>1
\end{array}
\right. ,
\end{eqnarray*}
for each bistrategy $(x,y)$ of the game. It is easy to realize that the two
strategies $a$ and $b$ are super-reactive for Emil, and, because they are
maxima of the set $E$ with respect to the preorder $\geq _{e}$, they are
indifferent. Let us see this directly. The set of Frances' strategies for
which $a$ is a possible reaction is $e^{-}(a)=F$, from which immediately
follows that $a$ is a maximum of the space $(E,\geq _{e})$ (no Emil's
strategy can be more reactive than $a$ inasmuch the strategy $a$ reacts to
all the Frances' strategies). Analogously, we can proceed for $b$ (for which
the situation is exactly the same).

\bigskip

\textbf{Remark (on the indifference in reactivity of strategies).} We note
that the reactivity indifference of two Emil's strategies $x$ and $x^{\prime
}$ is equivalent to the relation $e^{-}(x)=e^{-}(x^{\prime })$. In fact, the
preorder $\leq _{e}$ is induced by the function $e^{\leftarrow }$ of $E$ in $%
\mathcal{P}(F)$ defined by $x\mapsto e^{-}(x)$ with respect of the set
inclusion, and therefore $x$ and $x^{\prime }$ are equivalent in reactivity
if and only if they have the same value in $e^{\leftarrow }$.

\bigskip

\section{\textbf{Maximally reactive strategies}}

\bigskip

For the concept of maximal element in preordered spaces and its developments
we follow \cite{CA1}.

\bigskip

\textbf{Definition (of maximally reactive strategy).} \emph{Let }$(e,f)$ 
\emph{be a decision form game upon the underlying strategy pair }$(E,F)$%
\emph{. A Frances' strategy }$y\in F$ \emph{is called \textbf{maximally
reactive} if does not exist another Frances' strategy strictly more reactive
than }$y$\emph{\ (i.e., as we shall see later, if the strategy }$y$ \emph{is
not a sub-reactive strategy). In other terms, a Frances' strategy is called
maximally reactive if it is (Pareto) maximal in the preordered space }$%
(F,\geq _{f})$\emph{. Analogously, an Emil's strategy is called maximally
reactive if it is (Pareto) maximal in the preordered space }$(E,\geq _{e})$%
\emph{.}

\bigskip

\textbf{Example (of maximally reactive strategy). }Let $(e,f)$ be the
decision form game with strategy spaces $E=\left[ -1,2\right] $ and $
F=\left[ -1,1\right] $ and decision rules $e:F\rightarrow E$ and $%
f:E\rightarrow F$ defined by 
\begin{eqnarray*}
e(y) &=&\left\{ 
\begin{array}{cl}
\left\{ -1\right\}  & \mathrm{if}\;y<0 \\ 
E & \mathrm{if}\;y=0 \\ 
\left\{ 2\right\}  & \mathrm{if}\;y>0
\end{array}
\right. , \\
f(x) &=&\left\{ 
\begin{array}{cl}
\left\{ -1\right\}  & \mathrm{if}\;x<1 \\ 
F & \mathrm{if}\;x=1 \\ 
\left\{ 1\right\}  & \mathrm{if}\;x>1
\end{array}
\right. .
\end{eqnarray*}
The reciprocal correspondences of $e$ and $f$ are defined by 
\begin{eqnarray*}
e^{-}(x) &=&\left\{ 
\begin{array}{cl}
\left[ -1,0\right]  & \mathrm{if}\;x=-1 \\ 
\left\{ 0\right\}  & \mathrm{if}\;x\in \left] -1,2\right[  \\ 
\left[ 0,1\right]  & \mathrm{if}\;x=2
\end{array}
\right. , \\
f^{-}(y) &=&\left\{ 
\begin{array}{cl}
\left[ -1,1\right]  & \mathrm{if}\;y=-1 \\ 
\left\{ 1\right\}  & \mathrm{if}\;y\in \left] -1,1\right[  \\ 
\left[ 1,2\right]  & \mathrm{if}\;y=1
\end{array}
\right. .
\end{eqnarray*}
Hence we can easily note that the Emil's strategies $-1$ and $2$ are
maximally reactive. For instance, we shall study the strategy $2$. It is
sufficient to show that the subset $e^{-}(2)$ is not strictly included in
any other image $e^{-}(x)$, and this is evident. We have seen before that
these two maximal strategies are more reactive than all other Emil's
strategies $x\in \left] -1,2\right[ $, with respect to the rule $e$:
therefore all the Emil's strategies, with the exception of the two maximal
ones, are strictly less reactive than the maximal ones; moreover, all Emil's
strategies in $\left] -1,2\right[ $ are indifferent between them (they have
the same image through $e^{-}$), we see, so, that the interval $\left]
-1,2\right[ $ is even the set of all the minima of the preordered space $(E,%
\geq _{e})$. Analogously, we can prove that the strategies $-1$ and $1$ form
the maximal boundary of the preordered space $(F,\geq _{f})$.

\bigskip

\section{\textbf{Sub-reactive strategies}}

\bigskip

\textbf{Definition (of sub-reactive strategy).} \emph{A strategy }$s$\emph{\
of a player in a decision form game is said \textbf{sub-reactive} if there
exists a strategy }$s^{\prime }$\emph{\ of the same player strictly more
reactive than the strategy }$s$\emph{. In other terms, a Frances' strategy
is said sub-reactive if it is not (Pareto) maximal in the preordered space }$%
(F,\geq _{f})$\emph{. Analogously, an Emil's strategy is said sub-reactive
if it is not Pareto maximal in the preordered space }$(E,\geq _{e})$\emph{.}

\bigskip

\textbf{Example (of sub-reactive strategy). }Let $(e,f)$ be the game with
strategy spaces $E=\left[ -1,2\right] $ and $F=\left[ -1,1\right] $ and
decision rules $e:F\rightarrow E$ and $f:E\rightarrow F$ defined by 
\[
e(y)=\left\{ 
\begin{array}{cl}
\left\{ -1\right\}  & \mathrm{if}\;y<0 \\ 
E & \mathrm{if}\;y=0 \\ 
\left\{ 2\right\}  & \mathrm{if}\;y>0
\end{array}
\right. ,\;f(x)=\left\{ 
\begin{array}{cl}
\left\{ -1\right\}  & \mathrm{if}\;x<1 \\ 
F & \mathrm{if}\;x=1 \\ 
\left\{ 1\right\}  & \mathrm{if}\;x>1
\end{array}
\right. .
\]
We have seen before that the two Emil's maximal strategies $-1$ and $2$ are
more reactive than any other Emil's strategy $x\in \left] -1,2\right[ $,
with respect to the rule $e$: therefore all Emil's strategies, except the
maximal, are sub-reactive.

\bigskip

\section{\textbf{Elimination of sub-reactive strategies}}

\bigskip

\textbf{Definition (of reduced game by elimination of sub-reactive
strategies). }\emph{A game }$(e,f)$\emph{\ is said \textbf{reduced by
elimination of \emph{\textbf{sub-reactive} }strategies }if the maximal
(Pareto) boundaries of the preordered spaces }$(E,\geq _{e})$\emph{\ and }$%
(F,\geq _{f})$\emph{\ coincide\ with the strategy sets }$E$\emph{\ and }$F$%
\emph{, respectively.}

\bigskip

\textbf{Example (of not reduced game). }Let $(e,f)$ be the decision form
game with strategy spaces $E=\left[ -1,2\right] $ and $F=\left[ -1,1\right] $
and decision rules $e:F\rightarrow E$ and $f:E\rightarrow F$ defined by 
\begin{eqnarray*}
e(y) &=&\left\{ 
\begin{array}{cl}
\left\{ -1\right\}  & \mathrm{if}\;y<0 \\ 
E & \mathrm{if}\;y=0 \\ 
\left\{ 2\right\}  & \mathrm{if}\;y>0
\end{array}
\right. , \\
f(x) &=&\left\{ 
\begin{array}{cl}
\left\{ -1\right\}  & \mathrm{if}\;x<1 \\ 
F & \mathrm{if}\;x=1 \\ 
\left\{ 1\right\}  & \mathrm{if}\;x>1
\end{array}
\right. .
\end{eqnarray*}
The maximal boundaries of the preordered spaces $(E,\geq _{e})$ and $(F,\geq %
_{f})$ are the sets $\left\{ -1,2\right\} $ and $\left\{ -1,1\right\} $,
therefore the game is not reduced by elimination of sub-reactive strategies.

\bigskip

Before to proceed with the following definition, we recall the notion of
subgame of a decision form game.

\bigskip

\textbf{Definition (of subgame). }\emph{Let }$(e,f)$\emph{\ be a decision
form game upon the strategy pair }$(E,F)$\emph{\ and let }$(E^{\prime
},F^{\prime })$\emph{\ be a sub-strategy pair of }$(E,F)$\emph{, i.e. a pair
of subsets of }$E$\emph{\ and }$F$\emph{, respectively. We call \textbf{%
subgame of} }$(e,f)$\emph{\ \textbf{with underlying pair} }$(E^{\prime
},F^{\prime })$\emph{\ the pair of correspondence }$(e^{\prime },f^{\prime
}) $\emph{\ having as components the restrictions of the rules }$e$\emph{\
and }$f$\emph{\ to the pairs of sets }$(F^{\prime },E^{\prime })$\emph{\ and 
}$(E^{\prime },F^{\prime })$\emph{, respectively. We remember that, for
example, }$e^{\prime }$\emph{\ is the correspondence from }$F^{\prime }$%
\emph{\ into }$E^{\prime }$\emph{\ which sends a strategy }$y^{\prime }$%
\emph{\ of }$F^{\prime }$\emph{\ into the intersection }$e(y^{\prime })\cap
E^{\prime }$\emph{. In other terms, }$e^{\prime }$ \emph{sends every
strategy }$y^{\prime }$\emph{\ of }$F^{\prime }$\emph{\ into all Emil's
reaction strategies to }$y^{\prime }$\emph{\ which are in }$E^{\prime }$%
\emph{.}

\bigskip

\textbf{Definition (reduction of a game by elimination of sub-reactive
strategies). }\emph{Let }$G=(e,f)$\emph{\ be a decision form game with
underlying pair }$(E,F)$\emph{. We call \textbf{reduction of the game} }$%
(e,f)$\emph{\ \textbf{by elimination of\emph{\ sub-reactive }strategies} the
subgame }$(e^{\prime },f^{\prime })$\emph{\ of }$G$ \emph{with underlying
strategy pair the pair of the maximal Pareto boundaries }$\overline{\partial 
}_{e}E$\emph{\ and }$\overline{\partial }_{f}F$\emph{\ of the preorder
spaces }$(E,\geq _{e})$\emph{\ and }$(F,\geq _{f})$\emph{. In other terms,
the \textbf{reduction of the game} }$(e,f)$\emph{\ \textbf{by elimination of
the \emph{\textbf{sub-reactive} }strategies }is the game with decision rules
the restrictions }$e_{\mid (F^{\prime },E^{\prime })}$\emph{\ and }$f_{\mid
(E^{\prime },F^{\prime })}$\emph{, where }$E^{\prime }$\emph{\ and }$%
F^{\prime }$\emph{\ are the maximal Pareto boundaries }$\overline{\partial }
_{e}E$\emph{\ and }$\overline{\partial }_{f}F$\emph{\ of the preordered
spaces }$(E,\geq _{e})$\emph{\ and }$(F,\geq _{f})$\emph{.}

\bigskip

\textbf{Example (of reduction). }Let $(e,f)$ be the game with strategy
spaces $E=\left[ -1,2\right] $ and $F=\left[ -1,1\right] $ and decision
rules $e:F\rightarrow E$ and $f:E\rightarrow F$ defined by 
\begin{eqnarray*}
e(y) &=&\left\{ 
\begin{array}{cl}
\left\{ -1\right\}  & \mathrm{if}\;y<0 \\ 
E & \mathrm{if}\;y=0 \\ 
\left\{ 2\right\}  & \mathrm{if}\;y>0
\end{array}
\right. , \\
f(x) &=&\left\{ 
\begin{array}{cl}
\left\{ -1\right\}  & \mathrm{if}\;x<1 \\ 
F & \mathrm{if}\;x=1 \\ 
\left\{ 1\right\}  & \mathrm{if}\;x>1
\end{array}
\right. .
\end{eqnarray*}
The maximal boundaries of the preordered spaces $(E,\geq _{e})$ and $(F,\geq %
_{f})$ are the sets $E_{1}=\left\{ -1,2\right\} $ and $F_{1}=\left\{
-1,1\right\} $, therefore the game is not reduced, because they don't
coincide with the respective spaces. The reduction of the game $(e,f)$ by
elimination of sub-reactive strategies is the game with decision rules $%
e_{1}:F_{1}\rightarrow E_{1}$ and $f_{1}:E_{1}\rightarrow F_{1}$ defined by 
\begin{eqnarray*}
e_{1}(y) &=&\left\{ 
\begin{array}{cl}
-1 & \mathrm{if}\;y=-1 \\ 
2 & \mathrm{if}\;y=1
\end{array}
\right. , \\
f_{1}(x) &=&\left\{ 
\begin{array}{cl}
-1 & \mathrm{if}\;x=-1 \\ 
1 & \mathrm{if}\;x=2
\end{array}
\right. .
\end{eqnarray*}

\bigskip

\textbf{Example (of reduced game). }We note that the game $(e_{1},f_{1})$ of
previous example is reduced. In fact, the reciprocals correspondences of the
rules $e_{1}$ and $f_{1}$ are defined by 
\begin{eqnarray*}
e_{1}^{-}(x) &=&\left\{ 
\begin{array}{cl}
\left\{ -1\right\}  & \mathrm{if}\;x=-1 \\ 
\left\{ 1\right\}  & \mathrm{if}\;x=2
\end{array}
\right. , \\
f_{1}^{-}(y) &=&\left\{ 
\begin{array}{cl}
\left\{ -1\right\}  & \mathrm{if}\;y=-1 \\ 
\left\{ 2\right\}  & \mathrm{if}\;y=1
\end{array}
\right. .
\end{eqnarray*}
The maximal boundaries of the preordered spaces $(E_{1},\geq _{e_{1}})$ and $%
(F_{1},\geq _{f_{1}})$ are the sets $E_{2}=\left\{ -1,2\right\} $ and $
F_{2}=\left\{ -1,1\right\} $, respectively, therefore the game is reduced
because $E_{2}$ and $F_{2}$ coincide with the respective spaces. For an easy
determination of the two boundaries, we note that, for example, the
preordered space $(E_{1},\geq _{e_{1}})$ is isomorphic to the preordered
space with two elements $(\left\{ \left\{ 1\right\} ,\left\{ -1\right\}
\right\} ,\subseteq )$.

\bigskip

\section{\textbf{Iterated elimination of sub-reactivity}}

\bigskip

\textbf{Definition (of reducing sequence of a game). }\emph{Let }$%
G_{0}=(e_{0},f_{0})$\emph{\ be a game on a strategy base }$(E_{0},F_{0})$%
\emph{. We call \textbf{reducing sequence by elimination of\emph{\
sub-reactive }strategies of }}$G_{0}$\emph{\ the sequence of subgames }$%
G=(G_{k})_{k=0}^{\infty }$\emph{, with }$0$-\emph{term the game }$G_{0}$%
\emph{\ itself and with }$k$\emph{-th term the game }$G_{k}=(e_{k},f_{k})$%
\emph{, such that the strategy base }$(E_{k},F_{k})$\emph{\ of the game }$%
G_{k}$\emph{\ be the pair of maximal boundaries of the preordered spaces }$%
(E_{k-1},\geq _{e_{k-1}})$\emph{\ and }$(F_{k-1},\geq _{f_{k-1}})$\emph{, of
the }$(k-1)$\emph{-th subgame, for each positive integer }$k$\emph{. So, the
decision rules }$e_{k}$\emph{\ and }$f_{k}$\emph{\ are the restrictions to
the pairs }$(F_{k},E_{k})$\emph{\ and }$(E_{k},F_{k})$\emph{\ of the
decision rules }$e_{k-1}$\emph{\ and }$f_{k-1}$\emph{, respectively.}

\bigskip

\textbf{Definition (of solubility by iterated elimination of sub-reactive
strategies). }\emph{Let }$G_{0}=(e_{0},f_{0})$\emph{\ be a decision form
game, and let }$G$\emph{\ be its reducing sequence by elimination of
sub-reactive strategies. The game }$G_{0}$\emph{\ is called \textbf{solvable
by iterated elimination of \emph{sub-reactive strategies}} if there exists
only one bistrategy common to all subgames of the sequence }$G$\emph{. In
that case, that bistrategy is called the \textbf{solution \emph{by iterated
elimination of \emph{sub-reactive strategies of the game}}} }$G_{0}$\emph{.}

\bigskip

\textbf{Remark.} The definition of solubility by iterated elimination of
sub-reactive strategies is so equivalent to contain the intersection $%
\bigcap_{k=1}^{\infty }E_{k}\times F_{k}$ only one element.

\bigskip

\textbf{Remark.} If the game $G_{0}$ is finite, it is solvable by iterated
elimination of sub-reactive strategies if and only if there exists a subgame
of the sequence $G$ with only one bistrategy; in that case, that bistrategy
is the solution by iterated elimination of sub-reactive strategies of the
game $G_{0}$.

\bigskip

\section{\textbf{Relative super-reactivity}}

\bigskip

\textbf{Definition (of relatively super-reactive strategy).} \emph{Let }$%
(e,f)$\emph{\ be a two player decision form game. Let }$E^{\prime }$\emph{\
be a set of Emil's strategies to which Frances can react and let }$y_{0}$%
\emph{\ be a Frances' strategy. The strategy }$y_{0}$\emph{\ is called%
\textbf{\ relatively super-reactive for }}$E^{\prime }$\emph{\textbf{\ (with
respect to the decision rule }}$f$\emph{\textbf{)} if it is a possible
reaction to all the Emil's strategies in }$E^{\prime }$\emph{. In other
terms, a Frances' strategy }$y_{0}$\emph{\ is called relatively
super-reactive for }$E^{\prime }$\emph{\ if it belongs to the set }$f(x)$%
\emph{, for each Emil's strategy }$x$\emph{\ in }$E^{\prime }$\emph{.
Analogously, let }$F^{\prime }$\emph{\ be a set of Frances' strategies to
which Emil can react and }$x_{0}$\emph{\ an Emil's strategy. The strategy }$%
x_{0}$\emph{\ is called \textbf{relatively super-reactive for }}$F^{\prime }$%
\emph{\textbf{\ (with respect to the decision rule }}$e$\emph{\textbf{)} if
it is a possible reaction to all the Frances' strategies in }$F^{\prime }$%
\emph{. In other terms, an Emil's strategy }$x_{0}$\emph{\ is called
relatively super-reactive for }$F^{\prime }$\emph{\ if it belongs to the set 
}$e(y)$\emph{, for each Frances' strategy }$y$\emph{\ in }$F^{\prime }$\emph{%
.}

\bigskip

\textbf{Remark.} So the sets of Emil and Frances' relatively super-reactive
strategies for $F^{\prime }$ and for $E^{\prime }$ are the two intersections 
$\cap _{F^{\prime }}e=\bigcap_{y\in F^{\prime }}e(y)$ and $\cap _{E^{\prime
}}f=\bigcap_{x\in E^{\prime }}f(x)$. Evidently these intersections can be
empty.

\bigskip

\textbf{Example (of relatively super-reactive strategies). }Let $(e,f)$ be
the decision-form game with strategy spaces $E=\left[ -1,2\right] $ and $
F=\left[ -1,1\right] $ and decision rules $e:F\rightarrow E$ and $%
f:E\rightarrow F$ defined by 
\begin{eqnarray*}
e(y) &=&\left\{ 
\begin{array}{cl}
-1 & \mathrm{if}\;y<0 \\ 
E & \mathrm{if}\;y=0 \\ 
2 & \mathrm{if}\;y>0
\end{array}
\right. , \\
f(x) &=&\left\{ 
\begin{array}{cl}
-1 & \mathrm{if}\;x<1 \\ 
F & \mathrm{if}\;x=1 \\ 
1 & \mathrm{if}\;x>1
\end{array}
\right. .
\end{eqnarray*}
Emil has only a relatively super-reactive strategy for the Frances'
nonnegative strategies and only one relatively super-reactive strategy for
the Frances' nonpositive strategies. Indeed, we have 
\[
\bigcap_{y\in \left[ 0,1\right] }e(y)=E\cap \left\{ 2\right\} =\left\{
2\right\} 
\]
and 
\[
\bigcap_{y\in \left[ -1,0\right] }e(y)=\left\{ -1\right\} \cap E=\left\{
-1\right\} .
\]
Frances is in a similar situation for the Emil's strategies greater or equal
to $1$ and for the Emil's strategies less or equal to $1$, in fact, we have 
\[
\bigcap_{x\in \left[ 1,2\right] }f(x)=F\cap \left\{ 1\right\} =\left\{
1\right\} 
\]
and 
\[
\bigcap_{x\in \left[ -1,1\right] }f(x)=F\cap \left\{ -1\right\} =\left\{
-1\right\} .
\]

\bigskip

The following theorem has an obvious proof.

\bigskip

\textbf{Theorem (on reactivity). }\emph{Let }$x$\emph{\ be an Emil's
strategy. Then, the greatest among the sets }$F^{\prime }$ \emph{of
Frances's strategies\ such that the strategy }$x$\emph{\ is relatively
super-reactive for\ is the reactivity of }$x$\emph{.}

\bigskip

\textbf{Example (of reactivity). }Let $(e,f)$ be the game of above example.
The reactivity of the Emil's strategy $2$ is the interval $\left[ 0,1\right] 
$, the reactivity of the Emil's strategy $-1$ is the interval $\left[
-1,0\right] $. Indeed, these intervals are the biggest sets to which the
above strategies can react, respectively.

\bigskip

\section{\textbf{Dominant strategies}}

\bigskip

For the definition of normal form game used in this paper see \cite{CA4},
for the theory of normal form games we follow \cite{AU1}, \cite{AU2}, \cite
{OW}, \cite{OS} and \cite{MY}.

\bigskip

\textbf{Definition (of dominant strategy).} \emph{Let }$(u_{1},\geq )$\emph{%
\ be an Emil's utility function on the bistrategy space }$E\times F$\emph{\
of a strategy pair }$(E,F)$\emph{. An Emil's\ strategy }$x_{0}$\emph{\ in }$%
E $\emph{\ is said \textbf{dominant with respect to the \emph{utility }%
function }}$u_{1}$\emph{\ if, for each strategy }$x$\emph{\ in }$E$\emph{,
the inequality} 
\[
u_{1}(x_{0},y)\geq u_{1}(x,y), 
\]
\emph{holds, for each strategy }$y$\emph{\ in }$F$\emph{. In other terms, an
Emil's\ strategy }$x_{0}$\emph{\ in }$E$\emph{\ is said dominant if, for
each other strategy }$x$\emph{\ in }$E$\emph{, the function inequality} 
\[
u_{1}(x_{0},.)\geq u_{1}(x,.) 
\]
\emph{holds true. Analogously, let }$(u_{2},\geq )$\emph{\ be a Frances'
utility function on the bistrategy space }$E\times F$\emph{\ of a strategy
pair }$(E,F)$\emph{. A strategy }$y_{0}$\emph{\ in }$F$\emph{\ is said%
\textbf{\emph{\ \textbf{dominant with respect to the \emph{utility }function}%
} }}$u_{2}$\emph{\ if, for each }$y$\emph{\ in }$F$\emph{, the inequality} 
\[
u_{2}(x,y_{0})\geq u_{2}(x,y), 
\]
\emph{holds, for each strategy }$x$\emph{\ in }$E$\emph{. In other terms, a
Frances'\ strategy }$y_{0}$\emph{\ in }$F$\emph{\ is said }$u_{2}$\emph{%
-dominant if, for each other strategy }$y$\emph{\ in }$F$\emph{, the
function inequality} 
\[
u_{2}(.,y_{0})\geq u_{2}(.,y) 
\]
\emph{holds true.}

\bigskip

\section{\textbf{Dominant and super-reactive strategies}}

\bigskip

Let us see the first relationship between dominance and reactivity.

\bigskip

\textbf{Theorem (characterization of dominant strategies).} \emph{Let }$%
(u_{1},\geq )$\emph{\ and }$(u_{2},\geq )$\emph{\ be two Emil's and Frances'
utility functions, respectively, and let }$B_{1}$\emph{\ and }$B_{2}$\emph{\
be the respective best reply decision rules induced by the two functions }$%
u_{1}$\emph{\ and }$u_{2}$\emph{. Then, an Emil's strategy }$x_{0}$\emph{\
is }$u_{1}$\emph{-dominant if and only if it is }$B_{1}$\emph{-super
reactive and, analogously, a Frances' strategy }$y_{0}$\emph{\ is }$u_{2}$%
\emph{-dominant if and only if it is }$B_{2}$\emph{-super reactive.}

\emph{\bigskip }

\emph{Proof.} Let $x_{0}$ be a super-reactive strategy with respect to the
decision rule $B_{1}$. Then, the strategy $x_{0}$ belongs to the reaction
set $B_{1}(y)$, for each $y$ in $F$. So, for each $y$ in $F$, we have the
equality 
\[
u_{1}(x_{0},y)= \max u_{1}(.,y),
\]
that means 
\[
u_{1}(x_{0},y)\geq u_{1}(x,y),
\]
for each $x$ in $E$ and for each $y$ in $F$, that is the definition of
dominance. The vice versa can be proved by following the preceding steps in
opposite sense. $\blacksquare $

\bigskip

\section{\textbf{The preorder of dominance}}

\bigskip

\textbf{Definition (of dominance).} \emph{Let }$(u,\geq )$\emph{\ be a
normal-form game on the bistrategy space }$E\times F$\emph{\ of a strategy
base }$(E,F)$\emph{. We say that \textbf{an \emph{\textbf{Emil's} }strategy }%
}$x_{0}$\emph{\ \textbf{dominates (in wide sense)} \textbf{an other \emph{%
Emil's }strategy} }$x$\emph{\ \textbf{with respect to the \emph{\textbf{%
utility} }function }}$u_{1}$ \emph{if the partial function }$%
u_{1}(x_{0},\cdot )$\emph{\ is greater (in wide sense) of the partial
function }$u_{1}(x,\cdot )$\emph{. In this case we write }$x_{0}\geq
_{u_{1}}x$\emph{. We say that \textbf{an \emph{\textbf{Emil's} }strategy} }$%
x_{0}$\emph{\ \textbf{dominates strictly\ an other \emph{Emil's }strategy} }$%
x$\emph{\ \textbf{with respect to the \emph{\textbf{utility} }function} }$%
u_{1}$ \emph{if the partial function }$u_{1}(x_{0},\cdot )$\emph{\ is
strictly greater than the partial function }$u_{1}(x,\cdot )$\emph{. In this
case we write }$x_{0}>_{u_{1}}x$\emph{. We say that \textbf{an \emph{\textbf{%
Emil's} }strategy} }$x_{0}$\emph{\ \textbf{dominates strongly} \textbf{an
other \emph{Emil's }strategy} }$x$\emph{\ \textbf{with respect to the \emph{%
\textbf{utility} }function} }$u_{1}$ \emph{if the partial function }$%
u_{1}(x_{0},\cdot )$\emph{\ is strongly greater than the partial function }$%
u_{1}(x,\cdot )$\emph{. In that case we will write }$x_{0}\gg _{u_{1}}x$%
\emph{.}

\bigskip

\textbf{Memento (usual order on }$\mathcal{F}(X,\Bbb{R)}$\textbf{).} Let $X$
be a non-empty set, we remember that a real function $f:X\rightarrow \Bbb{R}$
is said greater (in wide sense) than an other function $g:X\rightarrow \Bbb{R%
}$, and we write $f\geq g$, if the wide inequality 
\[
f(x)\geq g(x), 
\]
holds for each $x$ in $X$. The above relation is said strict, and we will
write $f>g$, if the function $f$ is greater (in wide sense) than $g$ but
different. The $f$ is said strongly greater than $g$, and we write $f\gg g$,
if the strict inequality 
\[
f(x)>g(x), 
\]
holds true, for each $x$ in $X$. The majoring relation $\geq $ on the
function's space $\mathcal{F}(X,\Bbb{R)}$ is a order and it is called\emph{\
usual order of the space} $\mathcal{F}(X,\Bbb{R)}$. We note that the
relation $f\geq g$ is equivalent to the inequality 
\[
\inf (f-g)\geq 0. 
\]

\bigskip

\textbf{Remark.} We easily prove that the relation of dominance $\geq
_{u_{1}}$ is a preorder on $E$. Actually, it is the reciprocal image of the
usual order of the space of real functionals on $F$ (the space$\mathcal{\ F}
(F,\Bbb{R)}$) with respect to the application $E\rightarrow \mathcal{F}(F,%
\Bbb{R})$ defined by $x\mapsto u_{1}(x,.)$.

\bigskip

\textbf{Theorem (Characterization of the strict dominance for Weierstrass'
functions).} \emph{Let }$f_{1}:E\times F\rightarrow \Bbb{R}$\emph{\ be a
Weierstrass' functional (that is, assume that there are topologies }$\sigma $%
\emph{\ and }$\tau $\emph{\ on the sets }$E$\emph{\ and }$F$\emph{\
respectively such that the two topological spaces }$(E)_{\sigma }$\emph{\
and }$(F)_{\tau }$ \emph{are compact topological spaces and the function }$%
f_{1}$\emph{\ is continuous with respect to the product of those
topologies). Then, if the functional }$f_{1}$\emph{\ represents the Emil's
disutility, the condition }$x_{0}\gg _{f_{1}}x$\emph{\ is equivalent to the
inequality} 
\[
\sup (f_{1}(x_{0},.)-f_{1}(x,.))<0. 
\]

\emph{\bigskip }

\emph{Proof.} \emph{Necessity.} Let the strong dominance $x_{0}\gg _{f_{1}}x$
hold. Then the difference function $g=f_{1}(x_{0},.)-f_{1}(x,.)$ is negative
and moreover there exists (by the Weierstrass Theorem) a point $y_{0}$ in $F$
such that the real $g(y_{0})$ is the supremum of $g$, hence 
\[
\sup g=g(y_{0})<0. 
\]
\emph{Sufficiency} (the Weierstrass' hypothesis is not necessary). If the
supremum of $g$ is negative, every value of $g$ must be negative. $%
\blacksquare $

\bigskip

\section{\textbf{Dominance and reactivity}}

\bigskip

The following theorem explains the relationship between dominance and
reactivity comparison.

\bigskip

\textbf{Theorem (on the preorder of reactivity).} \emph{Let }$(u_{1},\geq )$%
\emph{\ and }$(u_{2},\geq )$\emph{\ be, respectively, two Emil's and
Frances' utility functions, and let }$B_{1}$\emph{\ and }$B_{2}$\emph{\ be
the best reply decision rules induced by the two functions }$u_{1}$\emph{\
and }$u_{2}$\emph{\ respectively. Then, the reactivity preorder }$\geq
_{B_{i}}$\emph{\ is a refinement of the preorder of dominance }$\geq
_{u_{i}} $\emph{.}

\emph{\bigskip }

\emph{Proof.} We shall show before that the preorder of reactivity refines
the preorder of dominance. Let $x_{0}\geq _{u_{1}}x$, then $%
u_{1}(x_{0},.)\geq u_{1}(x,.)$, from this functional inequality we deduce
that, if $y\in F$ and $x\in B_{1}(y)$ we have $x_{0}\in B_{1}(y)$. In fact, $%
x\in B_{1}(y)$ means that 
\[
u_{1}(x,y)= \max u_{1}(.,y) 
\]
but, because $u_{1}(x_{0},y)\geq u_{1}(x,y)$, we have also 
\[
u_{1}(x_{0},y)= \max u_{1}(.,y), 
\]
i.e., $x_{0}\in B_{1}(y)$. $\blacksquare $

\bigskip

The preorder of reactivity, in general, is a proper refinement of the
preorder of dominance, as the following example shows.

\bigskip

\textbf{Example.} Let $(B_{1},B_{2})$ be the Cournot decision form game with
bistrategy space $\left[ 0,1\right] ^{2}$ and net cost functions $f_{1}$ and 
$f_{2}$ defined by 
\[
f_{1}(x,y)=x\left( x+y-1\right) , 
\]
and, symmetrically, 
\[
f_{2}(x,y)=y\left( x+y-1\right) . 
\]
We easily see that every strategy in $\left[ 0,1/2\right] $ is strictly more
reactive than any strategy $x>1/2$, in fact the reactivity of any strategy $
x>1/2$ is the empty set (it is a non-reactive strategy). In particular, we
have $0>_{B_{1}}3/4$. On the other hand, the function $f_{1}(0,.)$ is the
zero real functional on $\left[ 0,1\right] $; on the contrary the partial
function $f_{1}(3/4,.)$ is defined by 
\[
f_{1}(3/4,.)(y)=(3/4)(y-1/4), 
\]
for each $y$ in $\left[ 0,1\right] $; since this last function has positive
and negative values, it is incomparable with the zero function, with respect
to usual order of the space of functions $\mathcal{F}(F,\Bbb{R})$.
Consequently, the preorder $\geq _{B_{1}}$ is a \emph{proper refinement} of
the preorder $\geq _{f_{1}}$.

\bigskip

\section{\textbf{Non-reactivity and strong dominance}}

\bigskip

Another concept used for normal-form games is that of strongly dominated
strategy (it is known in the literature also as strictly dominated strategy,
but we use this term for a less demanding concept).

\bigskip

\textbf{Definition (of strongly dominated strategy).} \emph{Let }$(u,\geq )$%
\emph{\ be a multi-utility function on the bistrategy space of a two player
game. Let }$(E,F)$\emph{\ be the pair of the strategy sets of the two
players (a game base). We say that a strategy }$x$\emph{\ in }$E$\emph{\ is 
\textbf{an \emph{\textbf{Emil's} }strongly dominated strategy }if there
exists an other strategy }$x^{\prime }$\emph{\ in }$E$\emph{\ such that the
strict inequality} 
\[
u_{1}(x,y)<u_{1}(x^{\prime },y), 
\]
\emph{holds for each strategy }$y$\emph{\ in }$F$\emph{. In other terms, we
say that a strategy }$x^{\prime }\in E$\emph{\ \textbf{strongly dominates} a
strategy }$x\in E$\emph{, and we write }$x^{\prime }\gg _{u_{1}}x$\emph{, if
the partial function }$u_{1}(x,\cdot )$\emph{\ is \textbf{strongly} less
then the partial function }$u_{1}(x^{\prime },\cdot )$\emph{.}

\bigskip

The following theorem explains the relationships between the non-reactive
strategies and the strongly dominated strategy.

\bigskip

\textbf{Theorem (strongly dominated strategies as never-best response
strategies).} \emph{Let }$(u_{1},\geq )$\emph{\ and }$(u_{2},\geq )$\emph{\
be respectively two Emil's and Frances' utility functions and let }$B_{1}$%
\emph{\ and }$B_{2}$\emph{\ be the best reply decision rules induced by the
two functions }$u_{1}$\emph{\ and }$u_{2}$\emph{\ respectively. Then, if a
strategy is strongly dominated with respect to the utility function }$u_{i}$%
\emph{\ it is non-reactive with respect to the decision rule }$B_{i}$\emph{.}

\bigskip

\emph{Proof.} Let $x_{0}$ be an Emil's $u_{1}$-strongly dominated strategy,
then there is at least a strategy $x$ in $E$ such that the inequality $%
u_{1}(x_{0},y)<u_{1}(x,y)$ holds true, for every $y$ in $F$. Hence the
strategy $x_{0}$ cannot be a best response to any strategy $y$ in $F$, since 
$x$ is a response to $y$ strictly better than $x_{0}$, for every $y$ in $F$;
so the reactivity of $x_{0}$, that is the set $B_{1}^{-}(x_{0})$, is empty. $%
\blacksquare $

\bigskip

To be a strongly dominated strategy is more restrictive than to be a never
best response strategy, as the following example shows.

\bigskip

\textbf{Example (an undominated and never-best response strategy).} Let $%
E=\left\{ 1,2,3\right\} $ and $F=\left\{ 1,2\right\} $ be the strategy sets
of a two player normal-form game $(u,\geq )$, and let $u_{1}$ be the Emil's
utility function defined by 
\begin{eqnarray*}
u_{1}(1,1) &=&u_{1}(1,2)=0 \\
u_{1}(2,1) &=&u_{1}(3,2)=1, \\
u_{1}(2,2) &=&u_{1}(3,1)=-1.
\end{eqnarray*}
We can summarize the function $u_{1}$ in a utility matrix $m_{1}$, as it
follows 
\[
m_{1}=\left( 
\begin{array}{cc}
0 & 0 \\ 
1 & -1 \\ 
-1 & 1
\end{array}
\right) .
\]
It is evident in the matrix $m_{1}$ that the Emil's strategy $1$ (leading to
the first payoff-row) is $u_{1}$-incomparable with the other strategies $2$
and $3$, and then it cannot be strongly dominated (neither strictly
dominated). On the other hand, $1$ is an Emil's never best response strategy
(i.e., it is non-reactive with respect to the best reply rule $B_{1}$).

\end{document}